\documentclass[fleqn,12pt,twoside]{article}
\usepackage{espcrc1}

\usepackage{epsfig}

\newcommand{\AmS}{{\protect\the\textfont2
  A\kern-.1667em\lower.5ex\hbox{M}\kern-.125emS}}

\newcommand{\lsim}{\,{\buildrel < \over {_\sim}}\,}
\newcommand{\gsim}{\,{\buildrel > \over {_\sim}}\,}

\title{Initial state of the QGP from perturbative QCD + saturation}

\author{Kari J. Eskola\address{Department of Physics, University of Jyv\"askyl\"a\\
 P.O. Box 35, FIN-40351 Jyv\"askyl\"a, Finland\\
{\em email: kari.eskola@phys.jyu.fi}
}%
        \thanks{also: Helsinki Institute of Physics,
P.O.Box 64, FIN-00014 University of Helsinki, Finland}}

\begin{document}

\maketitle

\begin{abstract}
{\small The production of the initial state of the QGP in very
high-energy $AA$ collisions is discussed within the framework of
perturbative QCD and saturation. The next-to-leading order computation
of the transverse energy of minijets is reviewed. Saturation of
parton production, conjectured to occur at a dynamically determinable
perturbative scale, leads to estimates of the initial densities. The
final state multiplicities are predicted by assuming an isentropic
hydrodynamical further evolution. Comparison with RHIC data is
shown.  }
\end{abstract}

\begin{flushright}
\vspace{-8.5cm}
hep-ph/0111223
\vspace{7.5cm}
\end{flushright}

\section{INTRODUCTION}

Particle production at the early stages, $\tau\sim1/p_T$, in
ultrarelativistic heavy ion collisions is dominated by processes of
large transverse momentum scales $p_T$.  The partonic multiplicities
$N_{AA}(p_T\ge p_0,\sqrt s,\Delta y)$ and transverse energies
$E_T^{AA}(p_0,\sqrt s,\Delta y)$ produced in processes above some
scale $p_0\gg\Lambda_{\rm QCD}$ into a rapidity window $\Delta y$ are
calculable \cite{BM87,KLL,EKL} in perturbative QCD (pQCD) through the
formalism of collinear factorization.  Subsequently, the number and
energy densities due to the pQCD quanta can be estimated.

Towards smaller transverse momenta, i.e. later in formation time, the
number of produced gluons is expected to grow to the extent that
non-linearities start to dominate: a new dynamically generated scale
appears, further growth in the number of gluons becomes inhibited, and
gluon production ``saturates''
\cite{BM87,GLR,MQ,McLV,KV,MUELLER,KOVCHEGOV,KHARZEEV,EKRT,EKT1,EKT2,EMW,PIRNER}.
For sufficiently high cms-energies and large nuclei, the saturation
scale becomes perturbative, $\sim 1...2$ GeV at RHIC...LHC in central
$AA$ collisions of $A\sim 200$ \cite{EKRT,KHARZEEV}.

Gluon saturation can be viewed to take place already in the initial
wave functions of the colliding nuclei \cite{GLR,MQ}, and gluon
production around the dominant saturation scale describable in terms
of classical fields \cite{McLV,KV,MUELLER,KOVCHEGOV,KHARZEEV}.  As 
non-perturbative gluon production cannot be computed
from truly first principles, one may equally well try to estimate the
bulk production of gluons in an $AA$ collision by using the pQCD
component alone but by extending the perturbative computation from
large-$p_T$ down to a dynamically determined saturation scale
\cite{EKRT,EKT1,EKT2}.  I shall focus on the latter, the pQCD+final-state
saturation approach, in the following. The initial-state saturation
and classical gluon fields are discussed by L. McLerran,
R. Venugopalan and D. Kharzeev in these proceedings. See also
ref. \cite{EMW} for self-screening in parton production, and
\cite{PIRNER} for other studies of particle production with saturation.

In the final-state
saturation approach \cite{EKRT}  the formation time of
the system can be estimated from the saturation scale $p_{\rm sat}$ as
$\tau_i=1/p_{\rm sat}$. This makes the computation of the initial
number and energy densities possible: identifying the spatial and the
momentum rapidities, and focussing on a volume element $\Delta
V_i\approx \pi R_A^2\tau_i\Delta y$ around $\eta=y=0$, we get the
average initial densities
\begin{equation}
n_i(\tau_i,z\sim 0)
=\frac{N_{AA}(p_{\rm sat},\sqrt s,\Delta y)}{\pi R_A^2\tau_i\Delta y},
\quad\quad\quad
\epsilon_i(\tau_i,z\sim 0)
=\frac{E_T^{AA}(p_{\rm sat},\sqrt s,\Delta y)}{\pi R_A^2\tau_i\Delta y}.
\label{init_densities}
\end{equation}
These densities serve as initial conditions for the further
evolution of the system. I shall first discuss how the quantities
$N_{AA}$ and $E_T^{AA}$ are computed in pQCD with the conjecture of  
saturation, and then return to the multiplicity predictions for RHIC 
and LHC.

\section{MINIJETS AND THEIR TRANSVERSE ENERGY FROM  PQCD}

In the leading twist approximation the semihard partonic collisions
with $p_T\sim 1...2$ GeV are independent of each other. For an $AA$
collision at an impact parameter {\bf b} the average number of
minijets produced into $\Delta y$ with $p_T\ge p_0$ and the
transverse energy carried by them can be computed as
\begin{eqnarray}
N_{AA}({\bf b},p_0,\sqrt s,\Delta y)&=& 
T_{AA}({\bf b})\sigma_{\rm pQCD}\langle N\rangle(p_0,\sqrt s,\Delta y,A)\\
E_T^{AA}({\bf b},p_0,\sqrt s,\Delta y)&=& 
T_{AA}({\bf b})\sigma_{\rm pQCD}\langle E_T\rangle(p_0,\sqrt s,\Delta y,A).
\label{ET_and_N}
\end{eqnarray}
The nuclear overlap function $T_{AA}({\bf b})=\int d^2{\bf
s}\,T_A({\bf s})T_A({\bf b-s})$ takes care of the collision
geometry. With the Woods-Saxon nuclear densities $T_{AA}(0)\approx
{A^2}/{\pi R_A^2}$ \cite{EKL}.  The first moment $\sigma_{\rm
pQCD}\langle E_T\rangle$ of the transverse energy distribution and
$\sigma_{\rm pQCD}\langle N\rangle$ of the number distribution of the
minijets in $\Delta y$, are the quantities computable in pQCD.

\subsection{Leading order}

In leading order (LO), the semihard partonic collisions are $2\rightarrow2$
processes. Their inclusive cross section can be obtained through collinear 
factorization, 
\begin{equation}
\frac{d\sigma^{AA\rightarrow kl+X}}{dp_T^2dy_1dy_2}
= \sum_{ij}\, x_1f_{i/A}(x,Q
)\, x_2f_{j/A}(x_2,Q)\,
\sum_{kl}\frac{d\hat\sigma}{d\hat t}^{ij\rightarrow kl}
\label{minijets}
\end{equation}
with the momentum fractions $x_{1,2}=\frac{p_T}{\sqrt s}({\rm e}^{\pm
y_1}+{\rm e}^{\pm y_2})$, the sub-cross sections $\frac
{d\hat\sigma}{d\hat t}^{ij\rightarrow kl}\hspace{-0.5cm} \sim
\alpha_s^2$ and the factorization/renormalization scales $Q\sim
p_T$. For the 1...2 GeV minijets at mid-rapidity ($y_1$ or $y_2 \sim
0$) at $\sqrt s=200...5500$ GeV of interest here, typically $x\sim
10^{-2}... 10^{-4}$. Due to the dominance of gluons at small values of
$x$, the gluons also clearly dominate the production of minijets.

Nuclear effects in Eq. (\ref{minijets}) are taken into account through
nuclear shadowing of parton distributions: $f_{i/A}(x,Q)\ne f_i(x,Q)$.  For this
purpose, we apply the EKS98 nuclear modifications \cite{EKS98} which
are based on a DGLAP analysis with data from deep inelastic $lA$
scattering and the Drell-Yan process and sum rules as constraints.

The transverse energy distribution of the minijets with $p_T\ge p_0$
and $y\in\Delta y$ is \cite{EKL}
\begin{equation}
\frac{d\sigma}{dE_T}\bigg|_{p_0,\Delta y}  = \int dp_T dy_1 dy_2\,
S_2^{E_T}(p_1,p_2)\,\frac{1}{2}\frac{d\sigma}{dp_Tdy_1dy_2}.
\end{equation}
The key element above is the ``measurement function'' for $E_T$ in $\Delta y$, 
\begin{equation}
S_2^{E_T}(p_1,p_2)\equiv\delta(E_T-[\epsilon_1+\epsilon_2]p_T)
\Theta(p_T\ge p_0),
\end{equation}
where $\epsilon_i\equiv \Theta(y_i\in \Delta y)$. This measurement
function on one hand keeps track of which partons fall in $\Delta y$ and
on the other hand ensures that the collisions are hard enough to allow for a
perturbative treatment. The first moment of the $E_T$ distribution
becomes then
\begin{equation}
\sigma_{\rm pQCD}\langle E_T\rangle (p_0,\sqrt s,\Delta y) \equiv \int dE_T
E_T\frac{d\sigma}{dE_T}\bigg|_{p_0,\Delta y} = \int_{p_0}dp_T\int_{\Delta y}dy_1\int
dy_2\, p_T \frac{d\sigma}{dp_Tdy_1dy_2}.
\end{equation}

Similarly, for the number distribution of the semihard minijets in
$\Delta y$, the measurement function is
$S_2^N(p_1,p_2)\equiv\delta(N-[\epsilon_1+\epsilon_2])\Theta(p_T\ge
p_0)$, and the first moment
\begin{equation}
\sigma_{\rm pQCD}\langle N\rangle (p_0,\sqrt s,\Delta y)\equiv \int dN N\frac{d\sigma}{dN}\bigg|_{p_0,\Delta y} =
\int_{p_0}dp_T\int_{\Delta y}dy_1\int dy_2\, \frac{d\sigma}{dp_Tdy_1dy_2}.
\end{equation}

\subsection{Next-to-leading order}

In next-to-leading order (NLO), where $\hat \sigma\sim \alpha_s^3$,
one must include both virtual corrections to $2\rightarrow2$ processes
and real emissions, i.e. $2\rightarrow3$ processes. The squared matrix
elements for the different subprocesses were computed in $4-2\epsilon$
dimensions first by R.K. Ellis and Sexton \cite{ES}. The subtraction
procedure of getting from the squared matrix elements in $4-2\epsilon$
dimensions to the physical (jet) cross sections was introduced by
S. Ellis, Kunszt and Soper \cite{EKS}. A detailed documentation can be
found in \cite{KS}.

For the $2\rightarrow3$ contributions with massless partons,
singularities arise when the momentum of an external leg becomes soft,
or when any two external particles become collinear.  In $4-2\epsilon$
dimensions, these singularities appear as terms $\sim 1/\epsilon$ and
$1/\epsilon^2$.  Similar terms arise also from the 1-loop 
graphs (interfered with the LO $2\rightarrow 2$). In the sum of the
different contributions, formed for computing a cross section of a
physical observable, the terms singular at the limit
$\epsilon\rightarrow0$ cancel, provided that one properly defines
\begin{itemize}
\item infra-red safe generalization of the measurement function $S_2$ 
to $S_3$ for the 3-particle final state: 
$S_3\rightarrow S_2$ when any final state  parton  becomes 
collinear with any other parton in the final or initial state,
or, when any final state parton becomes soft \cite{EKS}.
\item NLO PDFs  to absorb one $1/\epsilon$ singularity in the initial state
($\overline{\rm MS}$ scheme),
\item infra-red safe choice of the renormalization and factorization scales.
\end{itemize}

In the computation of the minijet $E_T$ production in NLO in $pp$ and
$AA$ collisions \cite{ET1,ET2}, we have (see also \cite{LEONIDOV})
adopted the subtraction method \cite{KS}.  Although the minijet
measurement functions differ from those for observable jets, the
computational procedure itself is similar to that of inclusive jet
production \cite{KS}, see \cite{ET2} for details. For e.g. the
$2\rightarrow3$ terms we need to numerically evaluate 6-dimensional
integrals with various kinematical cuts.

The infra-red safe measurement functions $S_2^{E_T}$ for the
$2\rightarrow2$ terms and $S_3^{E_T}$ for the $2\rightarrow3$ terms
are designed to answer the following question: How much $E_T$ is
carried into $\Delta y$ by minijets produced in partonic collisions
where at least an amount $2p_0$ of $E_T$ is produced? They are
\begin{eqnarray} 
S_2^{E_T}(p_1^{\mu},p_2^{\mu}) &=&  
\delta(E_T-[\epsilon_1p_{T1}+\epsilon_2p_{T2}])
\Theta(p_{T1}+p_{T2}\ge 2p_0)\cr
S_3^{E_T}(p_1^{\mu},p_2^{\mu},p_3^{\mu}) &=& 
\delta(E_T-[\epsilon_1p_{T1}+\epsilon_2p_{T2}+\epsilon_3p_{T3}])
\Theta(p_{T1}+p_{T2}+p_{T3}\ge 2p_0)
\end{eqnarray}
with $y_i$ and $p_{Ti}$ refering to the rapidities and transverse momenta
of the final state partons.

The semi-inclusive minijet $E_T$-distributions can then be defined in NLO 
as
\begin{eqnarray}
\frac{d\sigma}{dE_T}\bigg|_{p_0,\Delta y} 
&=& 
\int d[PS]_2 \frac{d\sigma^{2\rightarrow 2}}{d[PS]_2}S_2^{E_T}(p_1^{\mu},p_2^{\mu}) +
\int d[PS]_3 \frac{d\sigma^{2 \rightarrow 3}}{d[PS]_3}S_3^{E_T}(p_1^{\mu},p_2^{\mu},p_
3^{\mu}),
\label{dET}
\end{eqnarray}
where $d[PS]_2$ ($d[PS]_3$) is the 2(3)-particle phase space volume element.
The first term now includes both $\sim \alpha_s^2$ and $\sim \alpha_s^3$ contributions. The first $E_T$ moment becomes now 
\begin{eqnarray}
\sigma_{\rm pQCD}\langle E_T\rangle(p_0,\sqrt s,\Delta y) \equiv
\int_0^{\sqrt s} dE_T\,E_T \frac{d\sigma}{dE_T}\bigg|_{p_0,\Delta y} 
= \sigma \langle E_T\rangle_{\Delta y,p_0}^{2\rightarrow2}
+ \sigma \langle E_T\rangle_{\Delta y,p_0}^{2\rightarrow3},
\label{sET}
\end{eqnarray}
where 
\begin{eqnarray}
\sigma \langle E_T\rangle_{\Delta y,p_0}^{2\rightarrow2} &=&
\int d[PS]_2 \frac{d\sigma^{2\rightarrow 2}}{d[PS]_2}
\bigg[\epsilon_1+\epsilon_2\bigg]p_{T2
}
\Theta(p_{T2}\ge p_0)\\
\sigma \langle E_T\rangle_{\Delta y,p_0}^{2\rightarrow3} &=&
\int d[PS]_3 \frac{d\sigma^{2 \rightarrow 3}}{d[PS]_3}
\bigg[\epsilon_1p_{T1}+\epsilon_2p_{T2}+\epsilon_3p_{T3}\bigg]
\Theta(p_{T1}+p_{T2}+p_{T3}\ge 2p_0).
\label{sET3}
\end{eqnarray}
For the renormalization and factorization scales we use 
$Q = n(\sum_i p_{Ti})/2$ with $n\sim 1$. 

\begin{figure}[hbt]
\vspace{-1.5cm}
\centerline{\epsfxsize=9cm 
\epsfbox{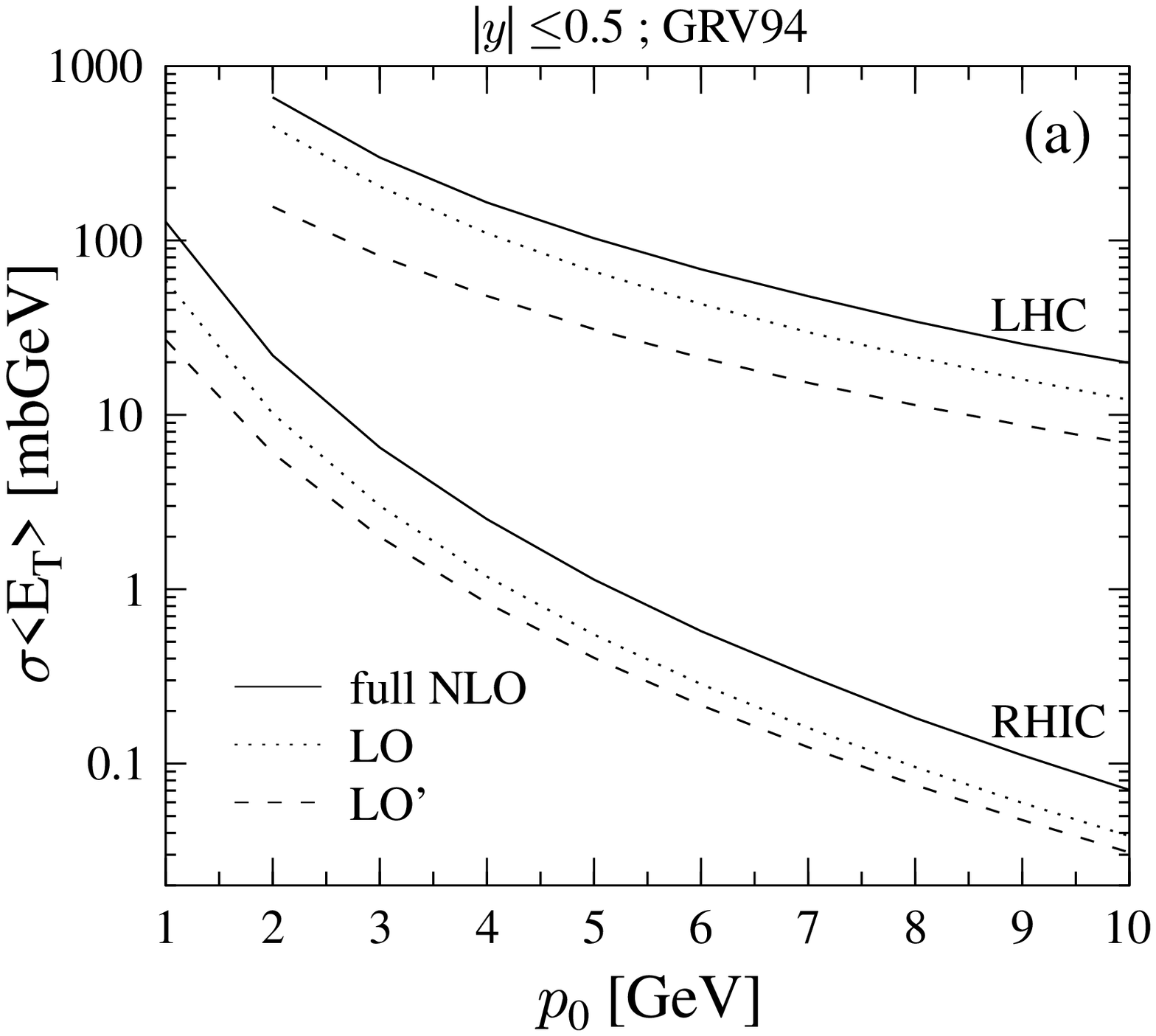}
\hspace{-1cm}
\epsfxsize=9cm 
\epsfbox{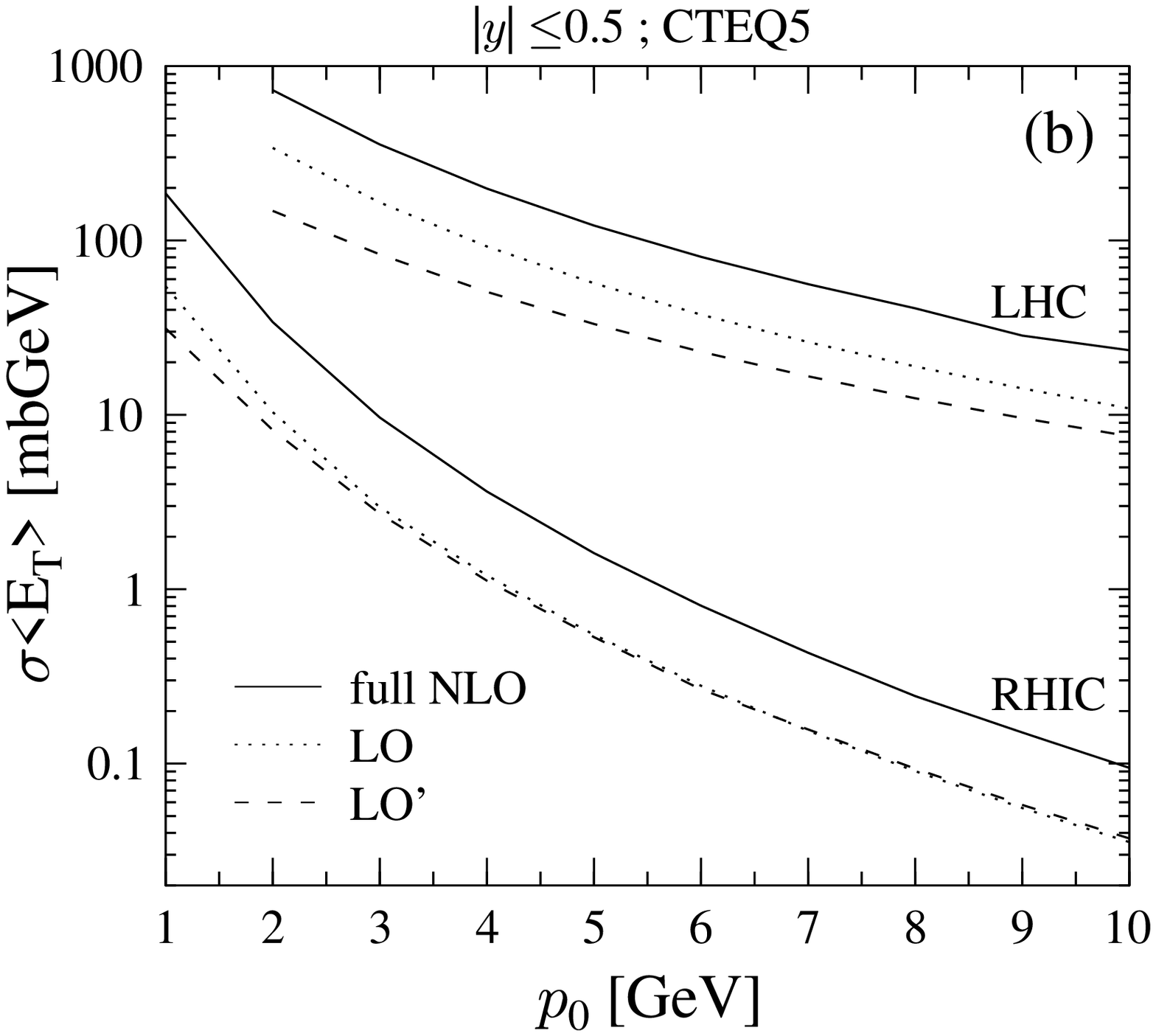}
}
\vspace{-2.5cm}
\caption{ \small $\sigma_{\rm pQCD}\langle E_T \rangle(p_0,\Delta y,\sqrt s) $ vs. $p_0$, from \cite{ET2}. For the details, see the text.
}
\label{sigma_ET}
\vspace{-0.5cm}
\end{figure}

Figure \ref{sigma_ET} shows the first moment of the minijet
$E_T$-distribution as a function of the minimum $p_T$-scale $p_0$,
computed for $pp$ collisions at $\sqrt s= 200$ GeV and $\sqrt s = 5500$
GeV. The solid curves are the full NLO result with 2-loop $\alpha_s$
and NLO PDFs, GRV94 \cite{GRV94} (left) and CTEQ5 (right)
\cite{CTEQ5}. The dotted curves (LO) show the lowest order
results ($\sim \alpha_s^2$) computed with the corresponding LO sets of
PDFs and 1-loop $\alpha_s$, and the dashed curves are the LO results
with NLO PDFs and 2-loop $\alpha_s$.

We see that even towards the perturbatively small scales of 1 GeV
at RHIC and 2 GeV at the LHC, the NLO results do not grow rapidly
relative to the LO (or LO') results. This speaks for the applicability
of pQCD at the semihard scales. If a $K$-factor is defined as NLO/LO,
we notice the stability of $K$ within $p_0=1...2$ GeV but that quite
obviously the $K$-factor depends on $\sqrt s$ and on the PDF set
used.  It is also interesting to observe that $\sigma\langle E_T\rangle$ 
at the LHC is insensitive to the choice of PDFs.

\begin{figure}[hbt]
\vspace{-1.0cm}
\centerline{\epsfxsize=9cm 
\epsfbox{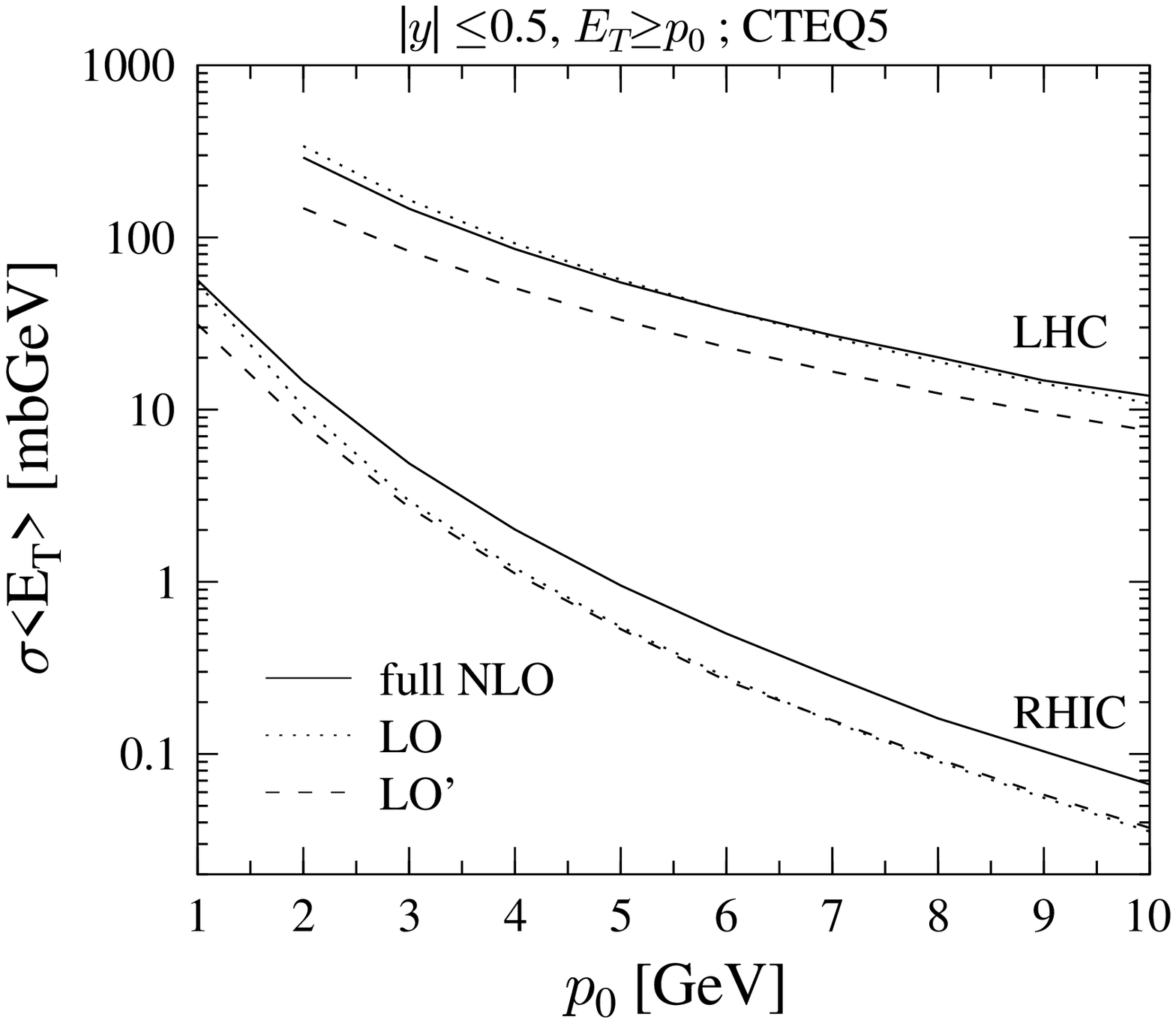}
\hspace{-1.cm}
\epsfxsize=9cm 
\epsfbox{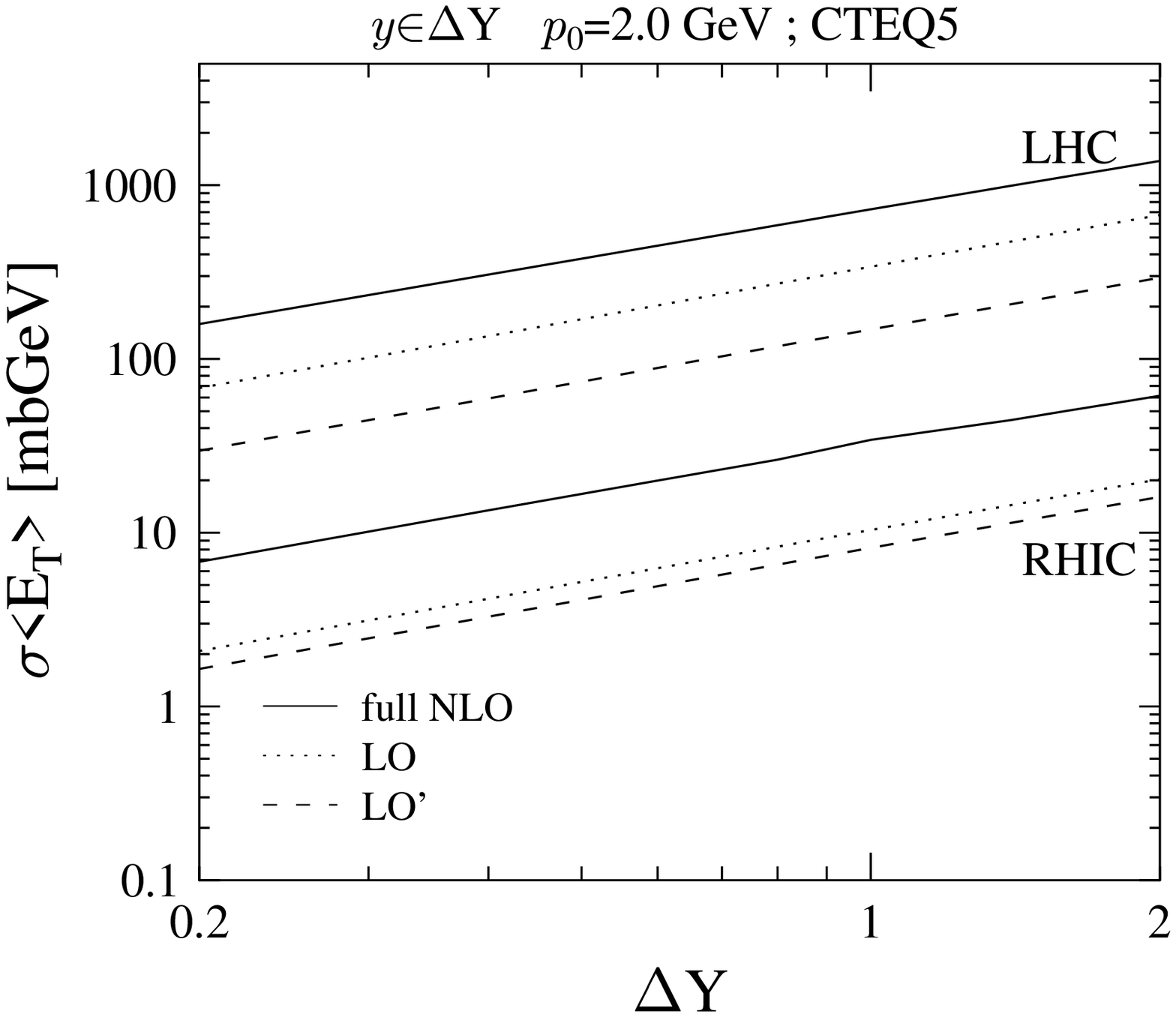}
}
\vspace{-2.5cm}
\caption {\small {\bf Left:} As Fig. 1 but with the region $E_T<p_0$ excluded.
{\bf Right:} The dependence of $\sigma\langle E_T\rangle$ on the rapidity interval $\Delta y$. The figures are from \cite{ET2}.
}
\vspace{-0.5cm}
\label{deltay_dep}

\end{figure}

Even though the NLO results are stable relative to the ones in LO, 
the $K$-factors are rather large, and one should be concerned of the 
convergence of the perturbation series. The dependence of 
$\sigma\langle E_T\rangle$ on the scale choice 
is very similar, quite weak, both in NLO and in LO \cite{ET2}.
The large NLO contributions mainly come from a kinematical domain $E_T<p_0$ 
which is not included at all in LO. The new region is due to processes 
where two hard partons fall outside $\Delta y$ and one parton with 
$p_T<p_0$ falls inside. To compare the NLO and LO results in the common 
kinematical region $E_T\ge p_0$, we have excluded the new region $E_T<p_0$
in Fig. \ref{deltay_dep} (left). The $K$-factors now reduce dramatically:
at the scales 1..2 GeV $K\sim$1 relative to LO.  This also speaks in favour 
of the applicability of pQCD at the semihard scales. At the same time it 
shows, however,  that to search for a convergence of the perturbative 
series, a NNLO computation is needed.

The inclusive jet cross sections are known to depend on the jet cone
radius $R$ as $d\sigma_{\rm jet}/dp_Tdy\sim CR^2+B\log R+D$
\cite{EKS}. If such a dependence of $\sigma\langle E_T\rangle$ on
$\Delta y$ were found, the initial densities in
Eqs. (\ref{init_densities}) would not be well-defined but the local
density would badly depend on $\Delta y$. Fig. \ref{deltay_dep} (right
panel) verifies that this is {\em not} the case but that the NLO
results for $\sigma\langle E_T\rangle$ depend in a same, linear, way
on $\Delta y$ as the LO results do. The initial energy density
$\epsilon_i$ at $\tau=1/p_0$ can thus be computed as in LO before but
using the $E_T$ obtained in NLO. Unfortunately, the NLO situation is
not as straightforward for the number density $n_i$. The number of
partons in $\Delta y$ is not an infrared-safe quantity unless one
introduces an additional resolution scale for two nearly collinear
partons.

To conclude the pQCD part, let me note that both $N_{AA}(p_0)$ (LO) and
$E_T^{AA}(p_0)$ (LO and NLO) depend strongly on the minimum
transverse momentum scale $p_0$.  Therefore we cannot compute the
initial densities $\epsilon_i$ and $n_i$ from pQCD alone but some
additional (phenomenological) element of QCD must be introduced
to dynamically determine the relevant scale $p_0$. In the following,
saturation of produced partons \cite{EKRT} is suggested as the 
mechanism which determines the dominant scale for gluon production.

\section{SATURATION IN PARTON PRODUCTION}

Saturation is a dynamical non-abelian feature of gluon fields.
Saturation phenomena were first discussed by Gribov, Levin and Ryskin
\cite{GLR} in the context of particle production in $pp$
collisions. Saturation in gluon production in $AA$ collisions was
first discussed by Blaizot and Mueller \cite{BM87}, then by McLerran and
Venugopalan in describing the particle production in $AA$ in terms of
classical gluon fields \cite{McLV,KV}, also studied by 
Mueller \cite{MUELLER}, Kovchegov \cite{KOVCHEGOV} and Kharzeev et al 
\cite{KHARZEEV}.  These studies
emphasize the saturation phenomena already in the initial wave
functions of the colliding objects.  Saturation of gluons at very
small values of $x$ then also saturates the actual production
(liberation) of gluons.

Saturation of gluon production can also be conjectured to arise in the
following way: Let us consider a nuclear disc of a transverse area $\pi
R_A^2$, filled with $N_g^A(Q)$ gluons, all produced at a scale $Q$ and
correlated over some longitudinal length.  The average area
density of gluons is $\rho(Q)\sim N_g^{A}(Q)/\pi R_A^2$.  The average
charge within some area $\pi R^2$ is then 
$\langle q_s \rangle \sim g_s\sqrt N \sim g_s \sqrt{\rho R^2}$, and 
the (Coulomb-like) potential $A\sim \langle q_s \rangle/R\sim g_s\sqrt \rho$. 
Saturation takes place when non-linearities in the field-strength 
tensor of QCD, $F\sim \partial A - g_s A^2$, become important, i.e. when 
$\partial A \,\lsim\, g_s A^2$. Dimensionally, $\partial A\sim
A/R\sim AQ$, and we obtain a criterion for saturation as $N_g^{A}(Q)\times
\alpha_s^2/Q^2\,\gsim\, \pi R_A^2$. Thus, if $N_g^A(Q)$ can be computed, 
the saturation scale $Q_{\rm sat}$ can be obtained. 

The above is, however, at best a scaling argument. In practise, 
in ref. \cite{EKRT}, we have used a geometrical saturation criterion
\begin{equation}
N_{AA}(p_0,\sqrt s,\Delta y)\times 
\frac {\pi}{p_0^2}=\pi R_A^2
\label{sat_crit}
\end{equation}
where we have neglected possible effects due the running of the
coupling, and have not included any group theoretical factors
explicitly. The correlation length in rapidity has been taken to be
$\Delta y=1$. In other words, we extend the computation of the number of 
minijets, $N_{AA}(p_0)$ discussed in the previous section, down to scale 
$p_0=p_{\rm sat}(\sqrt s,A)$, which is determined from Eq. (\ref{sat_crit}), 
and at which the produced gluons start to overlap transversally. The
multiplicity at saturation, $N_{AA}(p_{\rm sat})$, should then give a 
good first estimate of the total parton (gluon) production in central 
$AA$ collisions. Notice that it is the integral over $p_T$ that matters
and the contribution from $p_T\le p_{\rm sat}$ to the integral
is included through the saturation criterion.
\begin{figure}[t]
\vspace{-1.5cm}
\centerline{\epsfxsize=9cm 
\epsfbox{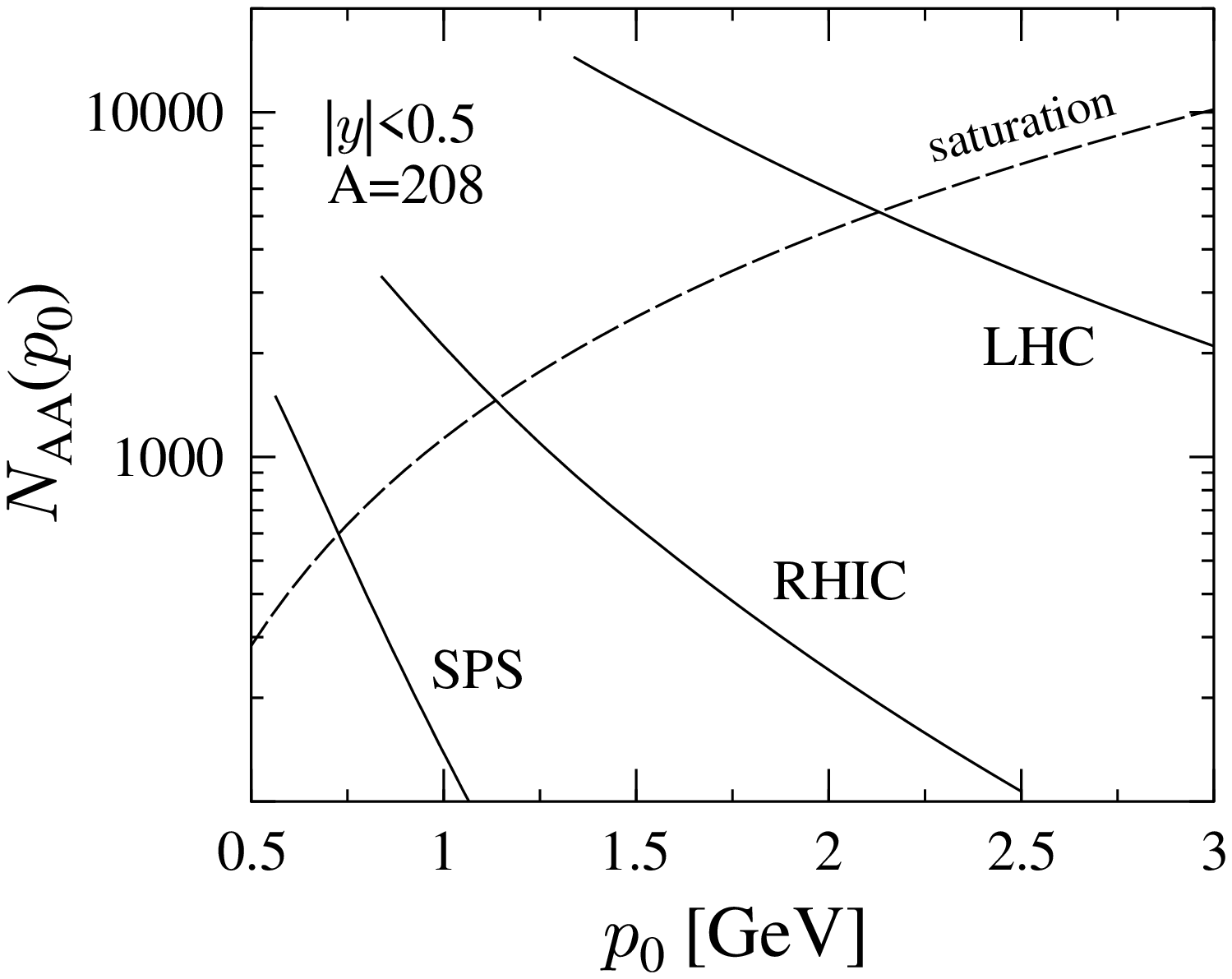}
\hspace{-1cm}
\epsfxsize=9cm\epsfbox{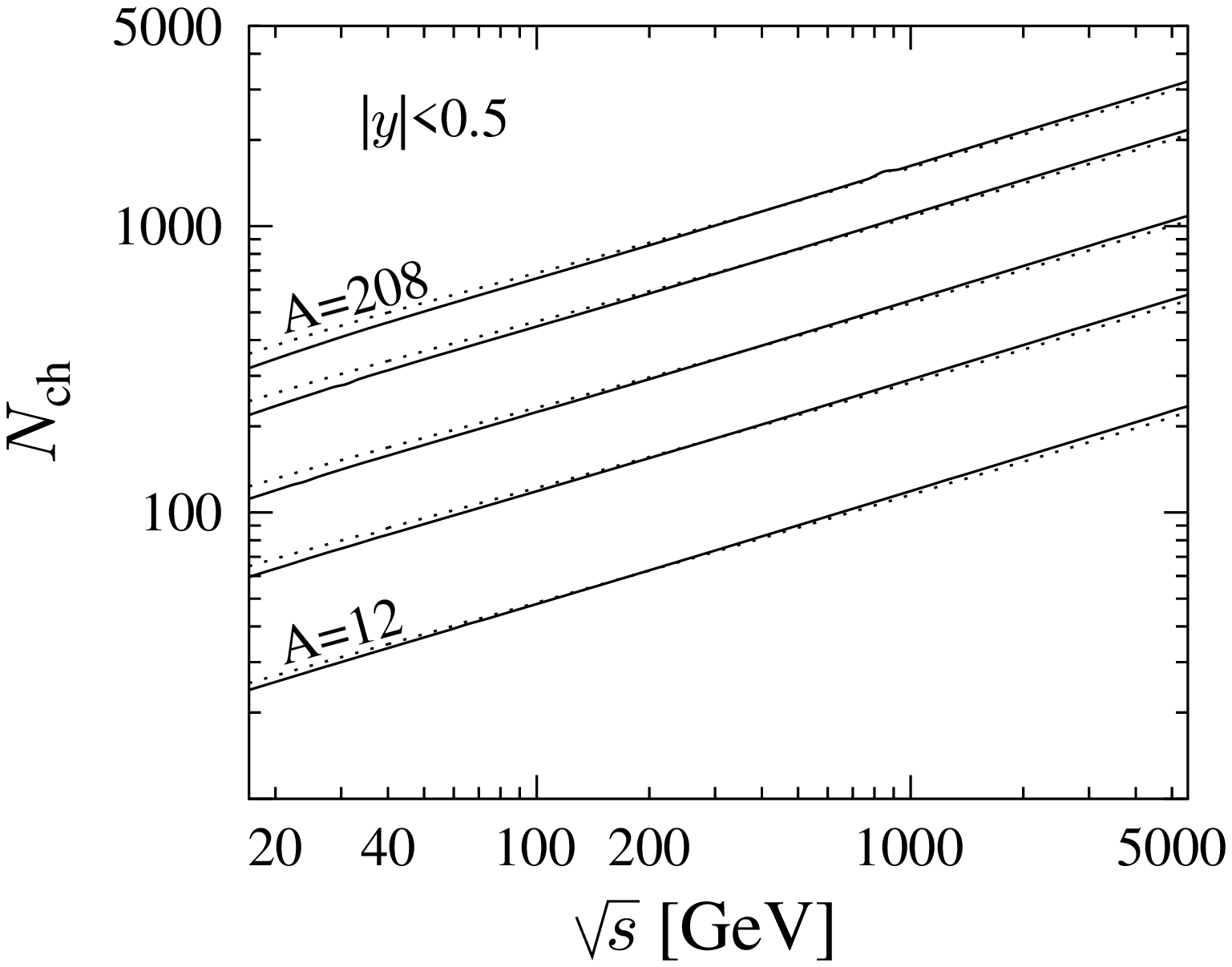}
}
\vspace{-2cm}
\caption {\small {\bf Left:} Determination of the saturation scale. 
{\bf Right:}  Charged particle multiplicity $dN_{\rm ch}/dy$ in central
$AA$ collisions for $A=208$, 136, 64, 32 and 12, as a function of $\sqrt s$. 
See the text for details. The figures are from \cite{EKRT}.
}
\vspace{-0.5cm}
\label{saturation}
\end{figure}

Figure \ref{saturation} (left) shows the average number partons
produced into $\Delta y=1$ with $p_T\ge p_0$ in a central $AA$
collision of $A=208$ at $\sqrt s/A=20$ GeV (SPS), 200 GeV (RHIC) at
5500 GeV (LHC). The curve labelled as ``saturation'' is $p_0^2R_A^2$
from Eq. (\ref{sat_crit}), and the saturation scale at each energy can
be read off from the intersection points of the curves.

It is quite interesting to notice that at the scaling limit
(neglecting the small-$x$ rise of the gluon densities, and the effects
of the phase space) $\sigma_{\rm pQCD}(p_0)\sim p_0^{-2}$, and the saturation
criterion (\ref{sat_crit}) results in $p_{\rm sat}\sim
A^{1/3}$ and in the multiplicity
 as $N_{AA}(p_{\rm
sat})\sim~A$ -- instead of the $A^{4/3}$ scaling typical for hard
processes.

With realistic gluon densities (GRV94LO \cite{GRV94}) including the
EKS98 shadowing effects \cite{EKS98}, and an overall $K=2$ to roughly
simulate the NLO effects, we have obtained in \cite{EKRT} the scaling
laws for $p_{\rm sat}(A,\sqrt s)$, $N_{AA}(p_{\rm sat},\Delta y)$ and
$E_T^{AA}(p_{\rm sat},\Delta y)$ presented in Table 1.  Recently, we
have also shown analytically how these scaling laws arise \cite{EKT2}
from the initial gluon densities, and that $N_{AA}(p_{\rm sat})\sim A
\alpha_s(p_{\rm sat}^2)xg_A(2p_{\rm sat}/\sqrt s,p_{\rm sat}^2)$,
where the explicit power of $\alpha_s$ now depends on the one included
explicitly in the saturation criterion (\ref{sat_crit}).  At the
collider energies, we see that the saturation scales are $p_{\rm
sat}\approx1\dots2\,{\rm GeV}\gg\Lambda_{\rm QCD}$ for the heavy
nuclei $A\sim200$.

\vspace{-0.0cm}
\begin{table}[htb]
\caption{\small The scaling laws for the saturation scale $p_{\rm sat}$,
initial multiplicity $N_{AA}(p_{\rm sat}, \Delta y)$ and transverse energy
$E_T^{AA}(p_{\rm sat},\Delta y)$, and the subsequent initial conditions 
for QGP  at $\tau_i$.}
\label{table:1}
\newcommand{\cc}[1]{\multicolumn{1}{c}{#1}}
\renewcommand{\tabcolsep}{1pc} 
\renewcommand{\arraystretch}{1.1} 
\begin{tabular}{@{}llll}
\hline
$A=208,	\Delta y=1$			& RHIC	& LHC & scaling in $A$ and $\sqrt s$\\
\hline
$\sqrt s/A$GeV           		& 200 	& 5500& \\
\hline
$p_{\rm sat}(A,\sqrt s)/$GeV               	& 1.13  & 2.13& $0.208A^{0.128}(\sqrt s)^{0.191}$ \\
$N_{AA}(p_{\rm sat},\Delta y)$		& 1440  & 5140&  $1.383 A^{0.922}(\sqrt s)^{0.383}$\\
$E_T^{AA}(p_{\rm sat},\Delta y)/$GeV	& 2360  & 17000& $0.386A^{1.043}(\sqrt s)^{0.595}$\\
$\tau_i=1/p_{\rm sat}$ (fm)   		& 0.17  & 0.093& \\
$n_i/{\rm fm}^{-3}$             	& 59.8  & 401&  $0.370A^{0.383}(\sqrt s)^{0.574}$ \\
$\epsilon_i/$GeVfm$^{-3}$       	& 98.2  &1330&  $0.103A^{0.504}(\sqrt s)^{0.786}$ \\
$T_i/$GeV (w. gluon d.o.f.)		& 0.62  & 1.19&  $0.111A^{0.126}(\sqrt s)^{0.197}$\\
\hline
\end{tabular}\\[2pt]
\vspace{-0.5cm}
\end{table}

The average initial conditions for the QGP obtained for central $AA$
collisions with $A=208$ for RHIC and LHC are also summarized in Table
1. The initial system is strongly gluon dominated, $\sim$ 90 \% of
partons produced at $p_{\rm sat}$ are gluons. We also note that from
the point of view of the computed average densities $\epsilon_i$ and
$n_i$, the system looks thermal: the energy per gluon is as in an ideal
gas, $E_T^{AA}(p_{\rm sat})/N_{AA}(p_{\rm sat})\approx 2.7 T_i$, where
$T_i$ is the temperature of an ideal massless gluon gas at the
computed energy density $\epsilon_i$. Therefore, it seems that for
thermalization gluon multiplication is not necessary, and early
thermalization, perhaps already at $\tau_i$, is possible (see also 
A. Mueller in these proceedings).

\section{FROM INITIAL TO FINAL MULTIPLICITIES}

In between the primary production
stage and decoupling, there is an expansion stage. A possible
description of this stage is given in terms of relativistic
hydrodynamics.  Encouraged by the initial conditions for the QGP
computed from the pQCD+saturation approach above, we assume
thermalization at $\tau_i=1/p_{\rm sat}$. The rapidity density of
entropy in a longitudinally boost-invariant system is $S_i\approx 3.6
N_{AA}(p_{\rm sat})$, counting only the gluonic degrees of freedom. In
an isentropic expansion entropy is conserved, so $S_i=S_f\approx
4N_f$, including only pions in the final state. To a first approximation, we 
can estimate the final state charged-particle multiplicity  in central 
$AA$ collisions directly 
from the initial conditions as
\begin{equation}
N_{\rm ch} \equiv dN_{\rm ch}/dy \approx \frac{2}{3} N_f = 
\left\{ \begin{array}{ll}

\frac{2}{3}1.24A^{0.922}(\sqrt s)^{0.383}
			& \mbox{if $S_i$ computed from $N_{AA}(p_{\rm sat})$ }
 \\\vspace{0.1cm}
\frac{2}{3}1.16A^{0.92}(\sqrt s)^{0.40}, 
			&\mbox{if $S_i$ computed from $E_T^{AA}(p_{\rm sat})$ }
\end{array}
\right.
\label{scaling_law}
\end{equation}
This prediction is shown in Fig. \ref{saturation} (right).  As seen in
the figure, since the system looks thermal from the very beginning,
the scaling laws (\ref{scaling_law}) obtained are very close to each
other. This illustrates the basic idea in multiplicity predictions
from pQCD+saturation.  For a more detailed study with $\sqrt
s$-dependent $K$-factors, transverse profiles, transverse expansion
effects, more detailed EoS, detailed treatment of decoupling, more
complete list of hadrons, resonance decays, computation of particle
spectra and centrality cuts, see refs. \cite{ERRT,KHHET}, and
V. Ruuskanen and U. Heinz in these proceedings.

\section{COMPARISON WITH RHIC DATA}

\begin{figure}[hbt]
\vspace{-2.5cm}
\centerline{\epsfxsize=10cm 
\epsfbox{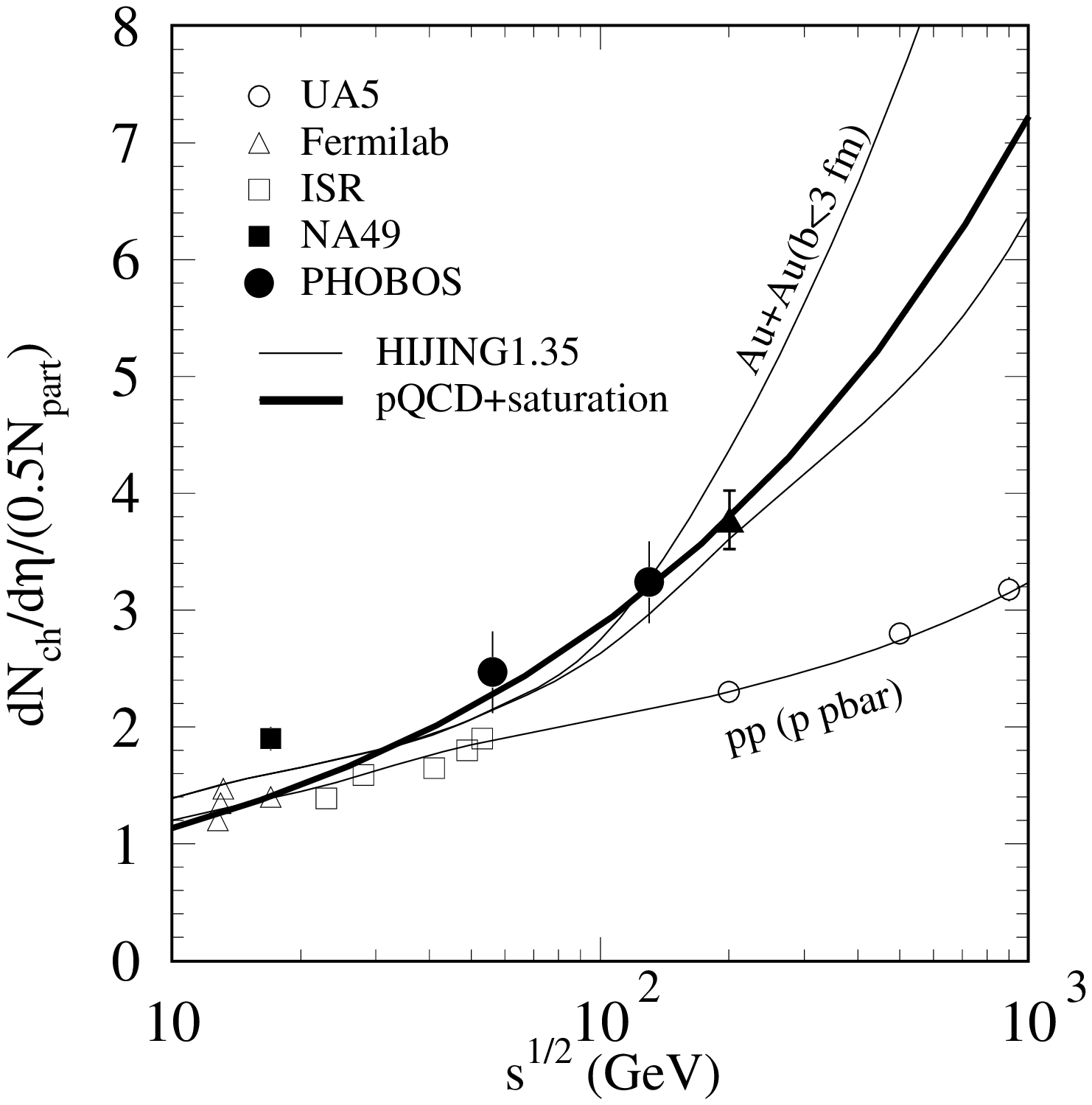}
\hspace{9cm}}
\vspace{-13.5cm}
\centerline{\hspace{8cm}
\epsfxsize=10cm\epsfbox{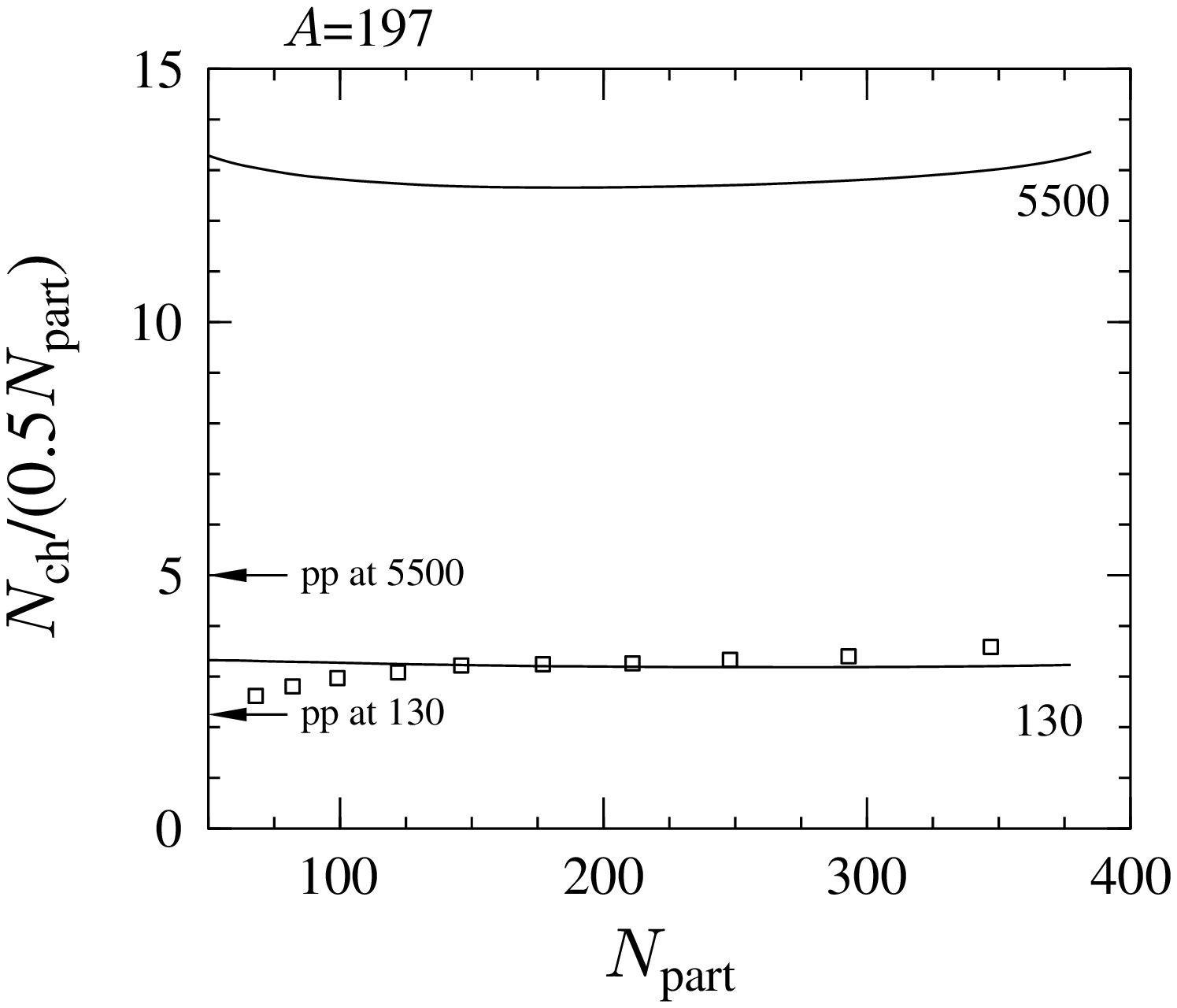} 
}
\vspace{-2.5cm}
\caption {\small {\bf Left:} $dN_{\rm ch}/d\eta/(0.5N_{\rm part})$ as
a function of $\sqrt s$. The prediction from pQCD+saturation approach
\cite{EKRT} is shown by the thick solid line. The HIJING1.35
\cite{HIJING} prediction is shown by the solid lines both for $pp$ and
for Au+Au with jet quenching (upper curve) and without (lower). The
PHOBOS data \cite{PHOBOS1} is shown by the filled circles.  The figure
is from \cite{WG01}, I have added the new PHOBOS data point at $\sqrt
s=200$ $A$GeV \cite{PHOBOS2} and emphasized the saturation curve. {\bf
Right:} The same quantity as a function of the number of participants
for $\sqrt s=130$ and 5500 $A$GeV as predicted by the pQCD+saturation
approach \cite{EKT2,EKT1}. The data is from PHENIX \cite{PHENIX}.  }
\vspace{-0.5cm}
\label{GW_fig}
\end{figure}

The above were predictions about one year before the first data from
RHIC.  The PHOBOS experiment measured the charged-particle
multiplicity in Au-Au collisions both at $\sqrt s=56$ $A$GeV and at
$\sqrt s=130$ $A$GeV in Summer 2000 \cite{PHOBOS1}.  Fig. \ref{GW_fig}
from \cite{WG01} shows the first PHOBOS data (6\% centrality cut)
by the filled circles, and the pQCD+saturation (EKRT) prediction
\cite{EKRT} by the thick solid line. The HIJING predictions \cite{HIJING}
with and without jet quenching are drawn with the solid lines. The
EKRT prediction shown is from Eq. (\ref{scaling_law}) with the number 
of participants $N_{\rm part}=2A$, and an approximate factor 0.9 
to account for the conversion $y\rightarrow\eta$ at $y\sim0$.

I have also added into the figure the latest measurement by PHOBOS at
$\sqrt s=200$ $A$GeV (filled triangle) \cite{PHOBOS2}.  Considering
the theoretical uncertainties, the data confirms the EKRT prediction
amazingly well both in absolute magnitude and in the scaling with
$\sqrt s$ in the RHIC energy regime. It should be emphasized that
once the effective constants in the saturation criterion
(\ref{sat_crit}) have been verified by comparison with the data at
some cms-energy, the pQCD+saturation approach gives a definite
prediction for the multiplicities at other energies, and also for
central collisions of other nuclei. As observed in Fig. \ref{GW_fig},
HIJING1.35 with a soft and a hard ($p_0=2$ GeV) particle production
components, predicts too large a multiplicity at $\sqrt s=200$
$A$GeV. For further discussion, see \cite{XNW}, and X.-N. Wang in 
these proceedings.

{\bf Centrality dependence} of the charged-particle multiplicity has
been suggested as a further challenge for the different models
\cite{WG01}.  In the pQCD+saturation approach the dependence of the
centrality of the $AA$ collision can be studied by making the
saturation criterion local in the transverse plane. Noticing in Eq. 
(\ref{sat_crit}) that $N_{AA}/\pi R_A^2$ is the average transverse
density of produced partons, the saturation criterion generalizes as
\cite{EKT1}
\begin{equation}
\frac{dN_{AA}}{d^2s}=
 T_A({\bf b}-{\bf s})T_A({\bf s})\sigma_{\rm pQCD}\langle N\rangle(p_0,\sqrt s,\Delta y,A)= p_0^2/\pi
\end{equation}
for an $AA$ collision at an impact parameter {\bf b}. The saturation
momentum needs to be determined at each transverse location {\bf s},
and the multiplicity of produced partons is then $N_{AA}(b)=\int
d^2{\bf s}p_{\rm sat}({\bf b},{\bf s},\sqrt s,A)$.  At sufficiently
large values of {\bf s} or {\bf b}, $p_{\rm sat}\sim \Lambda_{\rm
QCD}$, where we should not trust the pQCD calculation anymore. To
avoid this, we have considered particle production only from the
region where $p_{\rm sat}\ge 0.5$ GeV \cite{EKT1}. Clearly, the
saturation model \cite{EKT1} is applicable in the region where
dominantly $p_{\rm sat}\gg \Lambda_{\rm QCD}$.  

In Fig. \ref{GW_fig} (right), we show the comparison of the prediction
\cite{EKT1} against the PHENIX data \cite{PHENIX}. Also a prediction for
the LHC is shown.  The RHIC data suggests that the pQCD+saturation
model works for central and nearly central collisions. However, for
non-central collisions at $b\gsim R_A$ and $N_{\rm part}\gsim 200$,
one obviously moves outside the validity region of the model; the
dominant saturation momenta become too small, and particle production 
becomes overestimated.

I would like to emphasize, however, that although the centrality
dependence does provide constraints for different models, the effect
is still fairly modest, $\sim 20\dots30$\% in the data at $N_{\rm
part}\gsim 100$, and that in the absence of a direct measurement of
$N_{\rm part}$ the systematic errors can be large.  A more dramatic
effect is predicted from the $\sqrt s$-dependence of the
charged-particle multiplicity of central collisions: as seen in
Fig. \ref{GW_fig} (right), $N_{\rm ch}/0.5N_{\rm part}$ at the LHC
should be almost 4 times that at RHIC \cite{EKT2}. The $A$-scaling of
different models will be best tested by data from central $AA$
collisions with different $A$ instead of varying {\bf b} at fixed $A$.

{\small \vspace{0.1cm} \noindent {\bf Acknowledgements.} I thank
K. Kajantie, V. Ruuskanen and K. Tuominen for discussions, and the
Academy of Finland (grant No. 773101) for financial support.  }

\end{document}